\documentclass[superscriptaddress,aps,twocolumn,amsmath]{revtex4}

\renewcommand{\vec}{\boldsymbol}
\usepackage{graphicx}

\begin{document}
\title{Maximal entanglement of two spinor Bose-Einstein condensates} 
\author{Michael W. Jack}
\surname{Jack}
\affiliation{Department of Physics and Astronomy and Rice Quantum
Institute, Rice University, Houston, Texas 77251}
\affiliation{NTT Basic Research Laboratories, NTT Corporation, 3-1, Morinosato Wakamiya, Atsugi-shi, Kanagawa 243-0198, Japan}
\author{Makoto Yamashita}
\affiliation{NTT Basic Research Laboratories, NTT Corporation, 3-1, Morinosato Wakamiya, Atsugi-shi, Kanagawa 243-0198, Japan}
\date{\today}

\email{michael@atomcool.rice.edu}
\begin{abstract}
Starting with two weakly-coupled anti-ferromagnetic spinor condensates, we show that by changing the sign of the coefficient of the spin interaction, $U_{2}$, via an optically-induced Feshbach resonance one can create an entangled state consisting of two anti-correlated ferromagnetic condensates. This state is maximally entangled and a generalization of the Bell state from two anti-correlated spin$-1/2$ particles to two anti-correlated spin$-N/2$ atomic samples, where $N$ is the total number of atoms. 
\end{abstract}
\maketitle 
Quantum entanglement is the single most fundamental difference between quantum mechanics and classical mechanics\cite{bell87}. It is also the key ingredient that enables information transmission and processing beyond that possible classically \cite{nielsen00}. 
Quantum degenerate gases\cite{anderson95,dalfovo99}, due to their purity, weak environmental coupling and high experimental control, are a natural candidate to achieve such entangled states. In particular, with the experimental success of all optical trapping of a Bose-Einstein condensate \cite{stamper-kurn98},  it is now possible to explore atomic gases with spin degrees of freedom. There have already been a number of proposals  to create multi-particle entanglement between atoms with an internal degree of freedom \cite{pu00,duan00,sorensen01,ng03}.  
These proposals have concentrated on creating spin-squeezed states\cite{kitagawa93} or pairwise entanglement\cite{wang03}, in contrast, in this letter we wish to propose a method to produce a maximally entangled state of the spin degree of freedom of two spatially separated spinor condensates via an anti-ferromagnetic to ferromagnetic transition.

The original maximally entangled Bell state consists of a pair of anti-correlated spin$-1/2$ particles
\begin{equation}
|{\rm B}\rangle=\frac{1}{\sqrt{2}}\left(\left|\textstyle{\frac{1}{2}}\right\rangle_{1}\left|\textstyle{-\frac{1}{2}}\right\rangle_{2}-\left|\textstyle{-\frac{1}{2}}\right\rangle_{1}\left|\textstyle{\frac{1}{2}}\right\rangle_{2}\right).\label{bell}
\end{equation}
We can also straightforwardly generalize the above state to an arbitrary spin$-l$ particle. This generalized Bell state is also maximally entangled and has the form
\begin{equation}
|{\rm B}_{l}\rangle=\frac{1}{\sqrt{2l+1}}\sum_{m_{1}=-l}^{l}(-1)^{m_{1}}|m_{1}\rangle_{1}|-m_{1}\rangle_{2}\label{gb}
\end{equation}
such that each component of the spin, $m_{1}=-l,\cdots l$, of particle $1$ is anti-correlated with that of particle $2$.    The ``particles'' we consider in this paper are in fact two  ferromagnetic spin-$1$ atomic samples localized at the minimums of  a double-well optical potential, where the spin variable $l=N/2$ is equal to the number of atoms in each well and can have the mesoscopic values $\sim 10^{3}-10^{7}$\cite{dalfovo99}. Measurement of the entangled state is performed in a similar way to the Stern-Gerlach experiment, where a magnetic field gradient applied to one of the atomic samples will spread the angular momentum components spatially which can then be measured by absorption imaging\cite{dalfovo99}. This will project the sample into a definite component of the angular momentum and, due to the anti-correlation, the other sample will be projected into the opposite component.


The low energy collisions of spin-$1$ atoms trapped in an optical potential can be described by the interaction potential: 
\begin{equation}
\hat{V}(\vec{r}_{1}-\vec{r}_{2})=\frac{4\pi\hbar^{2}}{m}\delta(\vec{r}_{1}-\vec{r}_{2})\left[a_{0}\hat{P}_{0}+a_{2}\hat{P}_{2}\right],
\end{equation}
where $a_{F}$ is the scattering length in the total spin $F=0,2$ channel and $m$ is the atomic mass\cite{ho98}. Here $\hat{P}_{F}$ is a projection operator which projects the pair of atoms, $1$ and $2$, into a total hyperfine spin $F$ state. Following the initial proposal of Fedichev {\em et al.} \cite{fedichev96}, recent work has shown that the magnitude and the sign of the scattering lengths can  be tunned via all optical Feshbach resonances\cite{bohn97,fatemi00,thalhammer04,theis04}. The enhanced experimental control this provides opens up a new arena of possibilities for quantum state control in these spin systems, of which, the present work is one example. 

We consider the case of a symmetric double-well potential (such as that recently realized at MIT \cite{shin04}) where the system can be described in terms of well-localized spatial modes  in the two wells of the potential each containing $N/2$ atoms.  In addition, the spin dependence of the collisions is assumed to be small, $|a_{0}-a_{2}|\ll|a_{0}|$, so that we can assume the spatial modes are the same for each spin state \cite{ho98,yi02}.
Defining the quantities $c_{0}=4\pi\hbar^{2}(a_{0}+2a_{2})/3m$ and $c_{2}=4\pi\hbar^{2}(a_{2}-a_{0})/3m$ the Hamiltonian of the system has the form
\begin{eqnarray}
\hat{H} & =& \hat{H}_{\rm hop}+\hat{H}_{\rm coll}+\hat{H}_{\rm spin}\label{total}\\
\hat{H}_{\rm hop} & =& -J\sum_{\alpha=-1,0,1}\hat{a}_{1,\alpha}^{\dagger}\hat{a}_{2,\alpha}+\hat{a}_{2,\alpha}^{\dagger}\hat{a}_{1,\alpha}\\
\hat{H}_{\rm coll}&=&\frac{U_{0}}{2}\sum_{i=1,2}\hat{n}_{i}(\hat{n}_{i}-1)\label{collHamiltonian}\\
\hat{H}_{\rm spin}&=&\frac{U_{2}}{2}\sum_{i=1,2}\left[\hat{L}_{i}^{2}-2\hat{n}_{i}\right]\label{spinHamiltonian}
\end{eqnarray}
where $\hat{a}_{i,\alpha}$ is the annihilation operator for an atom with spin projection $\alpha$ in the $i$th well.
Here $J=-\int d^{3}\vec{r} \psi_{1}^{*}(\vec{r})[-\hbar^{2}\nabla^{2}/2m+V_{\rm pot}(\vec{r})]\psi_{2}(\vec{r})$ and $U_{F}=c_{F}\int d^{3}\vec{r}|\psi_{i}(\vec{r})|^{4}$  are given in terms of the spatial modes $\psi_{i}(\vec{r})$. For each well we have defined the number operators $\hat{n}_{i}=\sum_{\alpha}\hat{a}_{i,\alpha}^{\dagger}\hat{a}_{i,\alpha}$ and the spin vectors
$\hat{\vec{L}}_{j}=\hat{L}_{j,z}\vec{z}+\frac{1}{2}(\hat{L}_{j,+}+\hat{L}_{j,-})\vec{x}+\frac{i}{2}(\hat{L}_{j,+}-\hat{L}_{j,-})\vec{y}$,
 where $\vec{z}$, $\vec{x}$, and $\vec{y}$ are unit vectors in a Cartesian coordinate system and
\begin{eqnarray}
\hat{L}_{j,z}&=&\hat{a}^{\dagger}_{j,1}\hat{a}_{j,1}-\hat{a}^{\dagger}_{j,-1}\hat{a}_{j,-1},\\
\hat{L}_{j,-}& =& \sqrt{2}(\hat{a}^{\dagger}_{j,-1}\hat{a}_{j,0}+\hat{a}^{\dagger}_{j,0}\hat{a}_{j,1})
\end{eqnarray}
 and $\hat{L}_{j,+}=\hat{L}_{j,-}^{\dagger}$. The above ``tight-binding'' Hamiltonian is valid when the hopping can be treated as a weak perturbation to two independent wells, i.e. the hopping energy must be much smaller than the mode spacing in the independent wells \cite{milburn97}.

When $J=0$,  it is convenient to introduce the simultaneous eigenstates of $\hat{n}_{j}$, $\hat{L}^{2}_{j}$ and $\hat{L}_{j,z}$: 
\begin{equation}
|n_{j},l_{j},m_{j}\rangle\propto (\hat{L}_{j,-})^{l_{j}-m_{j}}(\hat{A}_{j}^{\dagger})^{s_{j}}(\hat{a}_{j,1}^{\dagger})^{l_{j}}|{\rm vac}\rangle,
\end{equation}
 where the spin-singlet creation operator is defined by $\hat{A}^{\dagger}_{i}=\hat{a}_{i,0}^{\dagger2}-2\hat{a}_{i,1}^{\dagger}\hat{a}^{\dagger}_{i,-1}$ \cite{koashi00}. The singlet number $s_{j}$ satisfies the relation $2s_{j}=n_{j}-l_{j}$. More generally, we can derive the  operator identity
\begin{equation}
\hat{n}_{i}(\hat{n}_{i}+1)=\hat{L}_{i}^{2}+\hat{A}_{i}^{\dagger}\hat{A}_{i}, \label{singlesite}
\end{equation}
 which has the interpretation: atoms either contribute to the magnitude of the spin angular momentum or they form singlets. It is also useful to define the simultaneous eigenstates of $\hat{n}_{j}$, $\hat{L}^{2}_{j}$, the total angular momentum $\hat{L}^{2}=\hat{L}_{1}^{2}+\hat{L}_{2}^{2}+2\hat{\vec{L}}_{1}\cdot \hat{\vec{L}}_{2}$ and $\hat{L}_{z}=\hat{L}_{1,z}+\hat{L}_{2,z}$: $|n_{1},l_{1},n_{2},l_{2};l,m\rangle$, which are related to $|n_{1},l_{1},m_{1}\rangle_{1}|n_{2},l_{2},m_{2}\rangle_{2}$ via the 
usual Clebsch-Gordan coefficients.

In the limit of vanishing hopping ($J=0$), we can use the fact that the $\hat{L}_{i}^{2}$ commute with the Hamiltonian to determine the ground state.  In the case of anti-ferromagnetic interactions ($U_{2}>0$), the ground state of each well is the state that minimizes  $\langle\hat{L}_{i}^{2}\rangle$ \cite{law98, ho98} which, from Eq.(\ref{singlesite}), implies that all atoms form singlets: $\langle \hat{A}_{i}^{\dagger}\hat{A}_{i}\rangle\approx\langle \hat{n}_{i}(\hat{n}_{i}+1)\rangle$.  In the simplest case when $N$ and $n_{1}=n_{2}=N/2$ are all even,  the ground state is given by the product state
\begin{equation}
|{\rm AF}\rangle=|l_{1}=0,m_{1}=0\rangle_{1}|l_{2}=0,m_{2}=0\rangle_{2}\label{antiferro}
\end{equation}
or $|{\rm AF}\rangle=|l_{1}=0,l_{2}=0;l=0,m=0\rangle$, where we have suppressed the $n_{i}$ labels for notational convenience. Using similar arguments we see that in the case of ferromagnetic interactions ($U_{2}<0$), the angular momentum in each well takes on its maximum value:  $\langle \hat{L}_{i}^{2}\rangle=\langle \hat{n}_{i}(\hat{n}_{i}+1)\rangle$. Note that in both the ferromagnetic and anti-ferromagnetic cases there is no correlation between the two wells.

We are now in a position to describe the formation of the maximally entangled state $|{\rm B}_{l}\rangle$ using simple conservation of angular momentum arguments. Assume our system has been prepared in the above anti-ferromagnetic ($U_{2}>0$) ground state such that $l=0$.  Now let us use an optically-induced Feshbach resonance to tune the scattering lengths $a_{0}$ and $a_{2}$ independently, such that $U_{2}$ is tunned adiabatically through zero  until $U_{2}<0$. It is possible to do this far from resonance so the atom loss due to spontaneous emission from excited states can be minimized\cite{fatemi00} (the experimental requirements of this transition are discussed in more detail below). Since the angular momentum is conserved in this process we have  $\langle \hat{\vec{L}}_{1}\cdot\hat{\vec{L}}_{2} \rangle =-\langle \hat{L}_{1}^{2}\rangle=-\langle \hat{L}_{2}^{2}\rangle$, but, now as we reduce the hopping coefficient, just as in the ferromagnetic case, the system must maximize the magnitude of spin in each well i.e. $\langle \hat{L}_{i}^{2}\rangle=\langle \hat{n}_{i}(\hat{n}_{i}+1)\rangle$. The only way to satisfy both requirements is for the spins to be maximally anti-correlated: $\langle\hat{\vec{L}}_{1}\cdot\hat{\vec{L}}_{2} \rangle =-\langle \hat{n}_{1}(\hat{n}_{1}+1)\rangle$. In the limit of vanishing hopping (for the simplest case of $N$ even), $n_{1}=n_{2}=N/2$ and  the ground state is given by
\begin{equation}
|{\rm B}_{N/2}\rangle=|l_{1}=N/2,l_{2}=N/2,l=0,m=0\rangle,\label{singlet}
\end{equation} 
which is the spin-$l$ singlet state.  In terms of the angular momentum states of each well this state has the remarkable form  given by Eq.(\ref{gb}) where we have suppressed the $n_{i}$ and $l_{i}$ labels. The two wells are maximally entangled as the reduced density matrix formed by performing a partial trace over one of the wells, $\rho={\rm ptr}\{|{\rm B}_{N/2}\rangle\langle {\rm B}_{N/2}|\}$, yields a completely mixed state with $N+1$ degenerate eigenvalues\cite{nielsen00}. Generation of entangled states in two-level systems via adiabatic crossings similar to the present method have been discussed in Ref.\cite{bell02}.

In order to create an entangled state in a real dynamical process we must provide a coupling between the initial state $|{\rm AF}\rangle$ and the final entangled state. Hopping between the wells is one possible candidate to provide this coupling. Let us consider  the hopping to be a weak perturbation to two independent wells (i.e. $J\ll |U_{2}|\ll U_{0}$). To first order the hopping does not change the number in each well and the effective hopping Hamiltonian is given by
\begin{equation}
\hat{H}_{\rm hop}^{\rm eff}=\hat{P} \hat{H}_{\rm hop}(E^{(0)}-\hat{H}_{\rm site})^{-1} \hat{H}_{\rm hop} \hat{P},
\end{equation}
 where $\hat{P}$ is a projection operator onto the subspace of fixed number at each site and, since $|U_{2}|\ll U_{0}$, $E^{(0)}\approx U_{0}N(N/2-1)/4$ is the energy of the ground state without hopping. A similar approach to treat hopping was taken in Ref.\cite{ng03} for the case of two species in a double-well potential. In contrast to their scheme, where an entangled state is created from a highly excited state (different species in each wells) via hopping between the wells, our approach involves keeping the system close to the ground state and thus avoids decoherence due to excitations.  After some straight-forward algebra (and dropping the constant terms) we arrive at the effective Hamiltonian
\begin{eqnarray}
\hat{H}_{\rm eff}&=&\frac{U_{2}}{2}\sum_{i=1,2} \hat{L}^{2}_{i}-2\frac{J^{2}}{U_{0}}\left(\hat{\vec{L}}_{1}\cdot \hat{\vec{L}}_{2}+\hat{G}^{\dagger}\hat{G}\right)\label{effective1}\\
&=&-U'_{2}\hat{\vec{L}}_{1}\cdot \hat{\vec{L}}_{2}-2\frac{J^{2}}{U_{0}}\hat{G}^{\dagger}\hat{G}\label{effective2}
\end{eqnarray}
where $\hat{G}^{\dagger}=\hat{a}^{\dagger}_{1,0}\hat{a}^{\dagger}_{2,0}-\hat{a}^{\dagger}_{1,1}\hat{a}^{\dagger}_{2,-1}-\hat{a}^{\dagger}_{1,-1}\hat{a}^{\dagger}_{2,1}$ (interestingly, this is the creation operator of a spin-1 Bell state) and $U_{2}'=U_{2}+2J^{2}/U_{0}$.  In the form given by Eq.(\ref{effective2}) we have dropped the term proportional to the total angular momentum as it is a constant and plays no role in the dynamics. Note that Eq.(\ref{effective1}) shows that the hopping gives rise to a spin-spin coupling between the wells, which should be contrasted with a spin-spin coupling originating from a (usually weak) dipole-dipole coupling \cite{pu01}.   The energy states of interest are the eigenstates of $\hat{H}_{\rm eff}$ in the $\hat{L}^{2}=0$, $\hat{n}_{j}=N/2$ subspace. When $J=0$, for the symmetric case, the ``bare'' eigenvalues and corresponding eigenstates  are $E_{l_{1}}=-U_{2}'l_{1}(l_{1}+1)$ and $|l_{1},l_{2}=l_{1},l=0,m=0\rangle$, where $l_{1}=0,2,\cdots, N/2$ for $N/2$ even. 
These eigenstates are the generalized Bell states given in Eq.(\ref{gb}) as we can write
\begin{eqnarray}
|{\rm B}_{l_{1}}\rangle &=&|l_{1},l_{2}=l_{1},l=0,m=0\rangle\\
&=&(2l_{1}+1)^{-1/2}\sum_{m=-l_{1}}^{l_{1}}(-1)^{m}|l_{1},m\rangle_{1}|l_{1},-m\rangle_{2},
\end{eqnarray} 
using the Clebsch-Gordan coefficients. For $l_{1}>0$ they are therefore  maximally entangled states. A measure of the entanglement of a pure state is given by the  entanglement entropy: $S=-{\rm tr}\{\rho\log_{2}\rho\}$\cite{nielsen00}. For the states $|B_{l_{1}}\rangle$ we find that the entanglement entropy scales as the log of $l_{1}$: $S=\log_{2}(2l_{1}+1)$. 

\begin{figure}
\begin{center}
 \includegraphics[scale=0.4]{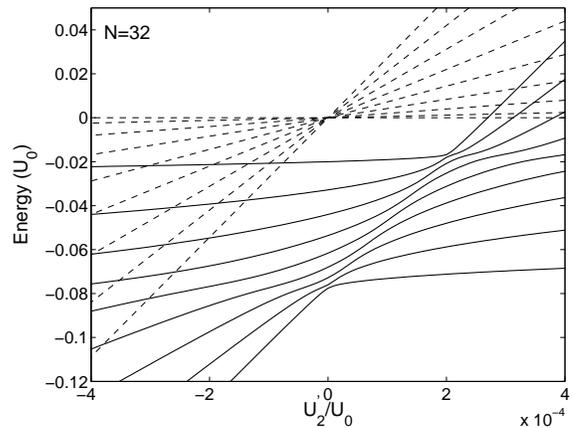}
\caption{\protect\label{levelCrossing} Example plot of the energy levels of the system as a function of $U_{2}'$. The thin dashed lines depict the $J=0$ case and the solid lines the $J\neq 0$ case. The energy offset between the two sets of energy levels has been altered for clarity. }
\end{center}
\end{figure}

Using the effective Hamiltonian (\ref{effective2}) we can analyze in detail the dynamical creation of an entangled state  from the initial state $|{\rm AF}\rangle$ in the case when $U'_{2}(t)$ is varied in time as  $U'_{2}(t)=-\alpha t$, where $\alpha>0$ is a constant. For $J=0$ the energies of the Hamiltonian (\ref{effective2}) undergo a multi-level crossing at $U'_{2}(t=0)=0$. For $J\neq 0$ the term proportional to $\hat{G}^{\dagger}\hat{G}$ couples these ``bare'' eigenstates and turns the level crossing into the avoided crossing shown in Fig.\ref{levelCrossing}.

\begin{figure}
\begin{center}
 \includegraphics[scale=0.4]{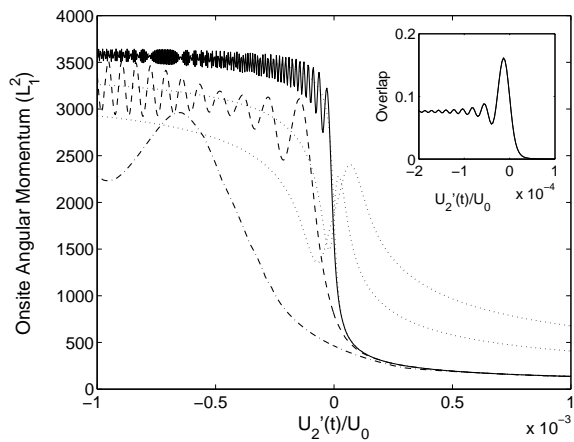}
\caption{\protect\label{timedep} Growth of the spin angular momentum in well $1$ as $U_{2}'(t)$ is varied from positive to negative from an initial anti-ferromagnetic state. In this plot $N=120$ and $J=U_{0}/20$. The solid line corresponds to $2\pi\Delta^{2}/(\hbar\alpha)=3200$, the dashed line to  $2\pi\Delta^{2}/(\hbar\alpha)=320$ and the dash-dotted line to $2\pi\Delta^{2}/(\hbar\alpha)=32$. The dotted lines show the angular momentum for the three lowest energy eigenstates. The inset shows the overlap of the time evolved state with the excited state for the $2\pi\Delta^{2}/(\hbar\alpha)=3200$ case. }
\end{center}
\end{figure}

Numerical diagonalization (up to $N=120$) of the effective Hamiltonian in the $| {\rm B}_{l_{1}}\rangle$ basis  shows that the minimum energy gap between the ground and first excited state occurs at $U'_{2}(t=0)=0$ and has the value $\Delta=16 J^{2}/U_{0}$ for $N\gg 1$.  Close to the adiabatic limit we expect that only the two lowest energy states play a role in the dynamics of the system and we can approximate the evolution as a Landau-Zener crossing \cite{zener32}. Numerical simulations (see Fig.\ref{timedep}) confirm that close to the adiabatic limit the probability of transitions to the first excited state follow the Landau-Zener law $P=e^{-2\pi\Gamma}$, where $\Gamma=\Delta^2/(\hbar dE/dt)$. Here $dE/dt$ is proportional to the slope of the difference between the ground and first excited state energy levels and has the form $dE/dt =\alpha f(N)$, where $f(N)$ is a function of $N$ and satisfies $f(N)\leq N(N/2+1)$ [c.f. Fig.\ref{levelCrossing}]. The adiabatic limit is then given by $\alpha\alt 2\Delta^2/[\hbar N(N+2)]$. Figure \ref{timedep} shows the evolution of $\langle \hat{L}_{1}^{2}\rangle(t)=-\langle \hat{\vec{L}}_{1}\cdot\hat{\vec{L}}_{2} \rangle(t)$ for different values of $\alpha$. From this figure we can see that for non-adiabatic changes the system tends to oscillate between the different eigenstates of the Hamiltonian (\ref{effective2}). 

As $m=m_{1}+m_{2}=0$, the states $|B_{l}\rangle$ are unaffected by a uniform magnetic field, however, a magnetic field difference between  the wells leads to a term of the form $H_{\rm field}\propto \hat{L}_{1,z}-\hat{L}_{2,z}$ in the Hamiltonian which will break the rotational symmetry and allow $\langle\hat{L}^{2}\rangle$ to vary in time. Spin flips and atom loss also give rise to changes in  $\langle\hat{L}^{2}\rangle$. In general, angular momentum conservation determines that a highly entangled state will form if the total spin angular momentum can be kept microscopic $(\langle \hat{L}^{2}\rangle \sim 1)$ during the time that $\langle L_{1}^{2}\rangle$ is increased from $\sim 1$ to $\sim N^2$. 

Once the entangled state is formed and the two wells are separated ($J\rightarrow 0$), applied magnetic fields will only lead to a phase rotation of the $|l,m\rangle_{1}|l,-m\rangle_{2}$ states but will not effect the level of entanglement between the two wells. Two other important decoherence mechanisms for the entangled state $|{\rm B}_{l}\rangle$ are spin flips and atom loss.
The state after a single spin flip in one of the wells is $|\psi_{\rm flip}\rangle\propto\hat{L}_{1,-}|{\rm B}_{l}\rangle$ which can be written as
\begin{equation}
|\psi^{l}_{\rm flip}\rangle\propto\sum_{m=-l+1}^{l}(-1)^{m}C^{l}_{m,-m+1}|l,m-1\rangle_{1}|l,-m\rangle_{2}\label{flip}
\end{equation}
where $C^{l}_{x,y}=\sqrt{(l+x)(l+y)}$. Since the angular momentum state of one well can be written as 
\begin{equation}
|n=l,l,m\rangle_{1}=\sqrt{\frac{(l+m)!}{(2l)!l!(l-m)!}}\hat{L}_{1,-}^{l-m}\hat{a}_{1,1}^{\dagger l}|{\rm vac}\rangle,
\end{equation}
 the annihilation operator acting on this state gives 
\begin{equation}
\hat{a}_{1,1} |l,l,m\rangle_{1}=\sqrt{\frac{(l+m)(l+m-1)}{2(2l-1)}}|l-1,l-1,m-1\rangle_{1}.
\end{equation}
 The state after the loss of one  atom from the entangled state is $|\psi^{l}_{\rm loss}\rangle\propto\hat{a}_{1,1}|{\rm B}_{l}\rangle$ which has the form
\begin{equation}
|\psi^{l}_{\rm loss}\rangle\propto\sum_{m=-l+2}^{l}(-1)^{m}C^{l}_{m,m-1}|l-1,m-1\rangle_{1}|l,-m\rangle_{2}\label{loss}
\end{equation}

\begin{figure}
\begin{center}
 \includegraphics[scale=0.3]{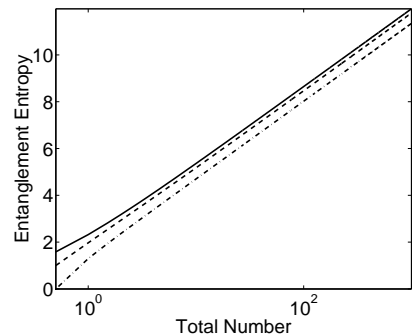}
\caption{\protect\label{decoherence} Effects of decoherence due to the loss of one atom and one spin flip on the entanglement as a function of total number. The solid line shows the entanglement entropy of the state $|{\rm B}_{N/2}\rangle$, the dashed line that corresponding to $|\psi^{N/2}_{\rm flip}\rangle$ and the dash-dotted line that corresponding to $|\psi^{N/2}_{\rm loss}\rangle$.}
\end{center}
\end{figure}

Since $\langle{\rm B}_{l}|\psi^{l}_{\rm flip}\rangle=\langle {\rm B}_{l}|\psi^{l}_{\rm loss}\rangle=0$, $|{\rm B}_{l}\rangle$ is not robust to spin flips or atom loss, as a  single spin flip or the loss of a single atom will transform it into an orthogonal state. However Eq.(\ref{flip}) and Eq.(\ref{loss}) show that the state after a spin flip or the loss of one atom is still highly entangled, and therefore can still be a useful source of entanglement for quantum teleportation\cite{nielsen00}, for example. 
In figure \ref{decoherence} we show the reduction of entanglement entropy after a spin flip or atom loss as a function of the total initial number of atoms. These results suggest that weak decoherence has only a small effect on the entangled state.  This is due to the fact that the decoherence effects on the level of the microscopic constituents (atoms) has little effect on the collective entanglement contained in the generalized Bell state. These results also demonstrate that even without exact symmetry between the wells (slightly different numbers of atoms in each well, for example), highly entangled states can still be created with this system. 

We now turn to a consideration of the general experimental requirements for using an optically-induced Feshbach resoance to tune the scattering lengths  such that $U_{2}\propto a_{2}-a_{0}$ changes sign and the interactions change from ferromagnetic to anti-ferromagnetic or vice versa\cite{ho98}. Optically-induced Feshbach resonances work by off-resonant photoassociation of two free atoms to an excited molecular state, thereby altering the scattering length of the atoms\cite{fedichev96}. Since, in general, the  excited molecular states with total spin $F=0$ and total spin $F=2$ are split by the hyperfine energy, $\hbar\omega_{\rm hf}=E_{2}-E_{0}$, (see, for example, \cite{abraham96})  it is in principle possible to tune the scattering lengths $a_{0}$ and $a_{2}$ independently as the photoassociative transitions to the excited molecular levels will experience different detunings.  Following Ref.\cite{bohn97} we can determine the change in the difference between the scattering lengths for a given detuning of the laser from resonance to be 
\begin{equation}
\Delta a= a_{2}-a_{0}+\frac{\Gamma}{2 k}\frac{\omega_{\rm hf}}{(E/\hbar-\Delta_{0})^{2}+(\gamma/2)^{2}-(\Gamma/2)^{2}},\label{deltaa}
\end{equation}
where $\Delta_{0}$ is the detuning of the laser from the $F=0$ molecular state, $E=\hbar^{2}k^{2}/2m$ is the atomic kinetic energy, $\gamma$ is the spontaneous emission rate and $\Gamma$ is the rate of photoassociation and is proportional to the product of the laser intensity and the overlap integral between the free atoms and the excited molecular state. This equation shows the ``leverage'' due to hyperfine splitting which enables us to tune the difference between the scattering lengths. 

Near an optical Feshbach resonance, spontaneous emission of molecules  in the excited molecular state inevitably leads to loss of atoms from the system.  In the present case this loss rate is given by the sum of losses due to the two molecular levels and has the form
\begin{equation}
K_{\rm inel}=\frac{4\pi\hbar}{m k}\frac{\gamma\Gamma}{(E/\hbar-\Delta_{0})^{2}+[(\Gamma+\gamma)/2]^{2}}.\label{loss}
\end{equation}
For simplicity, in Eq.(\ref{deltaa}) and (\ref{loss}), we have taken $\Delta_{0}\gg \omega_{\rm hf}$ and set the rates $\gamma$ and $\Gamma$ to be the same for the two excited molecular states. In contrast to the experiment described in \cite{theis04}, here we are not interested in large changes in the absolute value of the scattering length, but instead, in small relative changes in $a_{2}$ and $a_{0}$, which can be achieved far from the resonance with a high laser intensity thereby keeping the loss rate to a manageable level\cite{bohn97}. To illustrate this let us consider the specific case of $^{7}$Li (which is ferromagnetic with the scattering lengths $a_{0}=12$ a.u. and  $a_{2}=5$ a.u.\cite{mcalexander00}) and a photoassociative transition from the atomic continuum of the ground molecular potential $^{3}\Sigma_{u}^{+}$ to the $\nu=72$ vibrational level of the excited molecular potential $1^{3}\Sigma_{g}^{+}$, as discussed in Ref.\cite{bohn97}.   The hyperfine splitting between the $F=0$ and $F=2$ state in the excited molecular potential is $\omega_{\rm hf}=274.5$ MHz in this case\cite{abraham96}. For a laser intensity of $I=5$ kW/cm$^{2}$ and temperature of $1\mu$ K, $\Delta a$ becomes positive (and the interactions become anti-ferromagnetic as required) for detunings $\Delta<2.5$ GHz to the red of the $F=0$ level. At this detuning, losses occur at the rate of $K_{\rm inel}= 4 \times 10^{-12}$ cm$^{3}$/s, which is of the order of the rate of other loss mechanisms and is therefore of an acceptable level.

In summary, we have shown that starting with two weakly-coupled anti-ferromagnetic Bose-Einstein condensates it is possible to  form a highly entangled state consisting of two anti-correlated ferromagnetic condensates  by changing the sign of $U_{2}$ via optically-induced Feshbach resonance, while conserving the total spin-angular-momentum. The entanglement becomes maximal for symmetric wells in the limit of adiabatic changes.  In addition to the theoretical significance of this entangled state (as a generalization of the Bell state) and its novel method of creation (via an anti-ferromagnetic to ferromagnetic transition) we have shown that the quantity of entanglement of this state is approximately preserved under spin flips and loss of small numbers of atoms and therefore could be a useful resource for quantum information applications. 

An interesting generalization of this idea is to an optical lattice with $\sim 1$ atoms in each of the sites and hopping between neighboring sites\cite{greiner02}. Due to the number dependence of the adiabatic criteria described for the double-well case above, an entangled state in the lattice case may in fact be more easily realized experimentally.   A detailed  analysis of the dynamics of this system will be the subject of future work.

The authors would like to thank F. Morikoshi and K. Shimizu for useful discussions and H. Pu for reading this manuscript.


\end{document}